\documentclass[journal]{IEEEtran}
\usepackage[
backend=biber,
style=ieee
]{biblatex}
\usepackage[pdftex]{graphicx}
\usepackage{url}
\usepackage{subcaption}
\usepackage{amsmath}
\usepackage{url}

\begin{document}
\title{Haptic Shared Control for \\ Dissipating Phantom Traffic Jams}

\author{Klaas~O.~Koerten,
        David.~A.~Abbink,
        and~Arkady~Zgonnikov
\thanks{Department of Cognitive Robotics, Faculty of Mechanical, Maritime and Materials Engineering, Delft University of Technology, Delft,
Netherlands, e-mail: a.zgonnikov@tudelft.nl}
\thanks{Manuscript received xxx; revised xxx.}}

\maketitle

\begin{abstract}
Traffic jams occurring on highways cause increased travel time as well as increased fuel consumption and collisions. Traffic jams without a clear cause, such as an on-ramp or an accident, are called phantom traffic jams and are said to make up 50\% of all traffic jams. They are the result of an unstable traffic flow caused by human driving behavior. Automating the longitudinal vehicle motion of only 5\% of all cars in the flow can dissipate phantom traffic jams. However, driving automation introduces safety issues when human drivers need to take over the control from the automation. We investigated whether phantom traffic jams can be dissolved using haptic shared control. This keeps humans in the loop and thus bypasses the problem of humans' limited capacity  to take over control, while benefiting from most advantages of automation. In an experiment with 24 participants in a driving simulator, we tested the effect of haptic shared control on the dynamics of traffic flow, and compared it with manual control and full automation. We also investigated the effect of two control types on participants' behavior during simulated silent automation failures. Results show that haptic shared control can help dissipating phantom traffic jams better than fully manual control but worse than full automation. We also found that haptic shared control reduces the occurrence of unsafe situations caused by silent automation failures compared to full automation. Our results suggest that haptic shared control can dissipate phantom traffic jams while preventing safety risks associated with full automation.
\end{abstract}


\IEEEpeerreviewmaketitle

\section{Introduction}
\IEEEPARstart{T}{raffic} jams have far-reaching consequences, such as longer travel times, more frequent collisions, and increased fuel consumption and emissions due to the stop-and-go behaviour of vehicles in traffic jams~\cite{mahmud2012possible,wu2019tracking}. In many cases, traffic jams can be traced to accidents, sharp curves, on-ramps or sudden lane changes of vehicles. However, up to fifty per cent of all traffic jams do not have a distinct cause~ \cite{goldmann2020economic}; these are often called \textit{phantom traffic jams}. Investigating possible ways to reduce the occurrence of phantom traffic jams can therefore help to curb enormous costs associated with traffic congestion. 

Vehicle density plays a crucial role in the formation of phantom traffic jams. Each road has a certain critical density for each velocity. If car density increases beyond this density, the flow becomes unstable, and the average speed of the vehicles drops below the speed limit~\cite{treiber2013traffic}. However, even though high density is a prerequisite for the formation of phantom traffic jams, it is not its cause. What triggers phantom traffic jams are the velocity oscillations inherent in the behaviour of human drivers. While in sparse traffic these oscillations do not have a major impact on the traffic flow, when the density becomes high enough, these oscillations can propagate and amplify, thereby causing a phantom traffic jam~\cite{lee2019phantom}. \textcite{sugiyama2008traffic} and \textcite{tadaki2013phase} have shown this phenomenon occurring on a single-lane ring road with no external causes. Initially, drivers manage to keep a constant speed, but after some time, oscillations start to happen, the flow eventually becomes unstable, and stop-and-go waves start to form. Furthermore, commercially available adaptive cruise control systems can amplify disturbances of traffic flow just as much as human drivers on their own do~\cite{gunter2020commercially}. Overall, the reviewed studies show that natural human driving behaviour, as well as driving automation, can lead to the occurrence and preservation of phantom traffic jams. 

Existing approaches to the problem of phantom traffic jams can be divided into two categories. \textit{Centralised solutions} use sensors in the infrastructure to identify phantom traffic jams and solve them using dynamic traffic signs to open up additional traffic lanes or change the speed limit. These solutions have been shown to stabilise dense traffic~\cite{hoogendoorn2013assessment}, but require substantial adaptations to existing infrastructure; in addition, dynamic speed limits might slow down traffic when this is unnecessary. \textit{Decentralised solutions} use automated vehicles as agents to stabilise traffic locally to dissipate phantom traffic jams. One type of decentralised solution lets multiple automated vehicles drive behind one another in a stable platoon~\cite{kim2015demonstration}. Another type stabilises a traffic flow with a small number of automated vehicles (typically no more than 10\%) equipped with cruise control specifically developed for this~\cite{kreidieh2018dissipating, stern2018dissipation, vcivcic2018traffic}. Such automation-based solutions only require small adaptations to vehicles and can work for realistic penetration rates of cars equipped with adaptive cruise control.

However, automating vehicles comes at a cost. When the longitudinal motion of a car is fully automated, the drivers function as supervisors instead of operators, i.e. the human is taken out of the loop~\cite{parasuraman1987human}. According to \textcite{bainbridge1983ironies}, when humans become supervisors, their skills as operators of the task decline over time. However, they still remain responsible for taking over the task when things get too complicated for the automation. Typical problems associated with automation are the vigilance decrement ~\cite{davies1982psychology}, where the drivers' attention declines over time, and overreliance, where the drivers put too much trust in the automation and do not take over when necessary~\cite{parasuraman1987human}. Other disadvantages are increased reaction times and decreased performance when the human needs to take over. Human drivers sometimes do need to take over, because adaptive cruise control systems may experience difficulty in tracking a leading vehicle~\cite{son2006solution}, or in identifying an approaching stationary queue \cite{nilsson1996safety}. However, \textcite{rudin2004behavioural} show that reaction times of drivers increase when they rely on adaptive cruise control. Together, these effects could result in unsafe driving situations, especially in the dense traffic situations in which phantom traffic jams occur. This represents a potential pitfall of traffic stabilisation solutions based on full automation of the driving task.

As long as driving automation is not perfect, human drivers need to be able to take over control at any time. This requires them to stay engaged in the driving task, but how can they avoid losing vigilance while performing a daunting task of supervising automation? Continuously sharing the control between the human operator and the automation ensures driver engagement while still benefiting from the advantages of automation. In a simulator study, \textcite{jiang2021robust} showed how a sharing algorithm dampens traffic waves on a circular ring road. This shared controller takes as input the average control input of the human and an automated feedback controller. Essentially this means that a car equipped with this shared controller is automated for 50\% which could still result in the aforementioned safety issues. Imagine for example the case where a human wants to press the brake pedal suddenly, but the automation does not, then the human driver can only brake 50\% which could cause the car to not slow down fast enough to prevent an accident.

\textit{Haptic} shared control, on the other hand, does allow for such human intervention by making both the human operator and the automation exert forces on a control surface. The position of this control surface then determines the control action~\cite{abbink2010neuromuscular}. Haptic shared control for lateral vehicle motion can significantly improve safety compared to full automation when a silent automation failure happens~\cite{flemisch2008cooperative}. Previous research has reported improved performance with haptic feedback on the accelerator pedal for car-following~\cite{mulder2008haptic}, for making drivers more compliant to speed limits~\cite{adell2008auditory} and for promoting a more eco-friendly driving style~\cite{azzi2011eco, jamson2013design}. 

These results suggest that haptic shared control could potentially be an efficient solution to managing phantom traffic jams in a decentralized way while avoiding the pitfalls of automation. However, the effect of haptic shared control on traffic jams has not yet been investigated. This research aims to address this gap in the literature. The main contribution of this paper is an experimental evaluation of the efficacy of haptic shared control for dissipating phantom traffic jams jointly with the robustness of human drivers' reaction to silent automation failures. We propose a novel haptic shared gas pedal controller that incorporates a previously proposed algorithm for dissipating phantom traffic jams. We then test its performance in a driving simulator experiment with 24 participants. As benchmarks, we use the manual and fully automated conditions.

\section{Haptic Shared Controller Design}
The proposed haptic shared controller includes software and hardware components. The software part calculates the haptic forces that the controller applies to the pedals. The hardware includes a physical interface through which a human operator interacts with the automation.

\subsection{Software Design}
To implement haptic shared control, we first calculate the target longitudinal speed for the tested scenario using the algorithm proposed by~\textcite{stern2018dissipation}. This algorithm calculates the target speed based on the bumper-to-bumper gap and velocity difference between the ego vehicle and the leading vehicle. The speed is then fed into the cruise controller of the ego vehicle. We calculate the target speed in the following way:\\

\begin{equation}
    v^\textrm{cmd}=
    \begin{cases}
    0 & \textrm{if} \quad \Delta x \leq \Delta x_1\\
    
    v \frac{\Delta x - \Delta x_1}{\Delta x_2 - \Delta x_1} & \textrm{if} \quad \Delta x_1 < \Delta x \leq \Delta x_2\\
    
    v+(U-v)\frac{\Delta x - \Delta x_2}{\Delta x_3 - \Delta x_2} & \textrm{if} \quad \Delta x_2 < \Delta x \leq \Delta x_3\\
    
    U & \textrm{if} \quad \Delta x_3 < \Delta x
    
    \end{cases}
    \label{eq:vcmd}
\end{equation}
where 
\vspace{0.5cm}

\begin{equation}
    \Delta x_k= \Delta x_k^0 + \frac{1}{2d_k} (\Delta v \_)^2, \quad \textrm{for}\quad  k= 1,2,3
\end{equation}

In these equations, $v^\textrm{cmd}$ is the target speed, $U$ is the maximum speed on the road, $\Delta x$ is the bumper-to-bumper gap between the ego and leading vehicle, $v$ is the current speed, and $\Delta v\_$ is the difference in longitudinal velocity between the ego and leading vehicle. $\Delta x_k^0$ and $d_k$ are constant parameters, directly taken from \textcite{stern2018dissipation}: $\Delta x_1^0= 4.5m$, $\Delta x_2^0= 5.25m$, $\Delta x_2^0= 6.0m$, $d_1= 1.5\frac{m}{s^2}$, $d_2= 1\frac{m}{s^2}$ and $d_3= 0.5\frac{m}{s^2}$. 

As equation \ref{eq:vcmd} shows, the target speed is determined based on the gap to the leading vehicle, $\Delta x$. Different threshold for this gap, $\Delta x_k$, are defined as a combination of the base distance gap, $\Delta x_k^0$ and a term based on the velocity difference between the two vehicles, $\Delta v\_$. \textcite{stern2018dissipation} provide a more detailed description of this algorithm.

The difference between $v^\textrm{cmd}$ and $v$ is the input for the pedal controller. This controller does the following:
\
\begin{align}
    \textrm{if} \quad  v^\textrm{cmd}-v > 0  &\Rightarrow \textrm{accelerate}\nonumber\\
    \textrm{if} \quad 0 \geq v^\textrm{cmd}-v \geq -0.25 &\Rightarrow \textrm{do nothing}\nonumber\\
    \textrm{if} \quad v^\textrm{cmd}-v < -0.25   &\Rightarrow  \textrm{brake}
\end{align}

Equation 3 shows that the control actions are either to \textbf{1)} do nothing, making the car brake on the engine, where both pedals get released, \textbf{2)} press the accelerator pedal and release the brake pedal to accelerate or \textbf{3)} press the brake pedal and release the accelerator pedal to decelerate. When acceleration or deceleration needs to happen, the pedal position of the accelerator or the brake pedal is determined by feeding the difference in speed into the accelerator or decelerator controller respectively. Both controllers are PID controllers that produce the target pedal positions, $S^{\textrm{target}}_{\textrm{acc}}$ and $S^{\textrm{target}}_{\textrm{brake}}$. The gains of the accelerator controller are $K_p=1$, $K_i=0.01$ and $K_d=0.05$ and the gains of the decelerator controller are $K_p=0.7$, $K_i=-0.04$ and $K_d=0.1$. The controllers were tuned by running the ring road simulation and feeding the outputted target pedal position as the current position into the simulated ego vehicle. The gains were then adjusted until wave dampening behaviour similar to \textcite{stern2018dissipation} was achieved.

The haptic accelerator and brake pedals that we use are control interfaces using the control loader technology. Software controls the virtual mass, damping, stiffness, and the forces that the pedals exert on the driver's foot. The default behaviour of both pedals is spring-damper behaviour.\\

\begin{figure}
    \centering
    \includegraphics[width=1\linewidth]{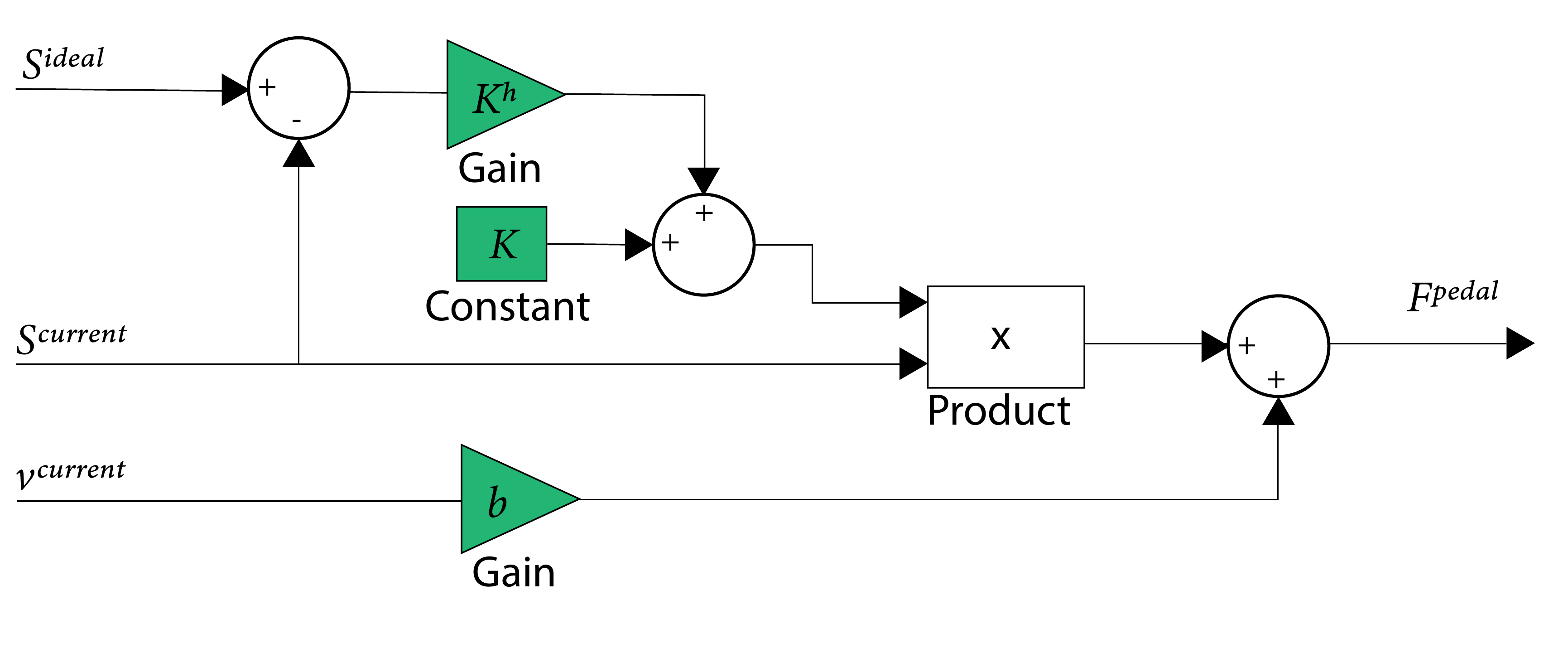}
    \caption{Block diagram of the accelerator pedal for the haptic shared control case}
    \label{fig:diagrampedal}
\end{figure}

For the haptic shared controller we use the difference between the current pedal position $S^{current}$ and the target position $S^{target}$ to set the stiffness $K$ of the accelerator pedal to $K^{hc}$ (Figure~\ref{fig:diagrampedal}):

\begin{equation}
    K^{hc}=K+K^{h}\times(S^p-S^{target})
\end{equation}

The choice to use stiffness feedback was made. This is done because then, when the pedal moves from the targeted position, it gets pushed back with a virtual spring that increases its force depending on the distance from the targeted position. This makes feedback less drastic than, for example, force feedback, that applies a sudden force when the pedal leaves the targeted position. This is also explained in \textcite{mulder2010design}. We initially provided feedback on both the accelerator and brake pedal. However, in pilot studies, participants reported that haptic feedback on the brake pedal did not make them feel safe in the car. This was because a braking action is done to quickly slow down the car. When the driver would want to press the brake sooner or later than the automation, the brake would be less stiff or stiffer than expected. This would result in unwanted behaviour of the brake pedal and therefore in an unwanted braking action, making the driver feel unsafe.

For these reasons, we made the choice to only provide haptic feedback on the accelerator pedal. Also, we set the increase in stiffness when the pedal needs to be released at $300 N/radian^2$  while the decrease in stiffness when the pedal needs to be pressed is to $30 N/radian^2$. We base this design choice also on pilot studies where the participants did not notice feedback when the gains were the same for increasing and decreasing the pedal stiffness.

\begin{figure}
    \centering
    \includegraphics[width=0.9\linewidth]{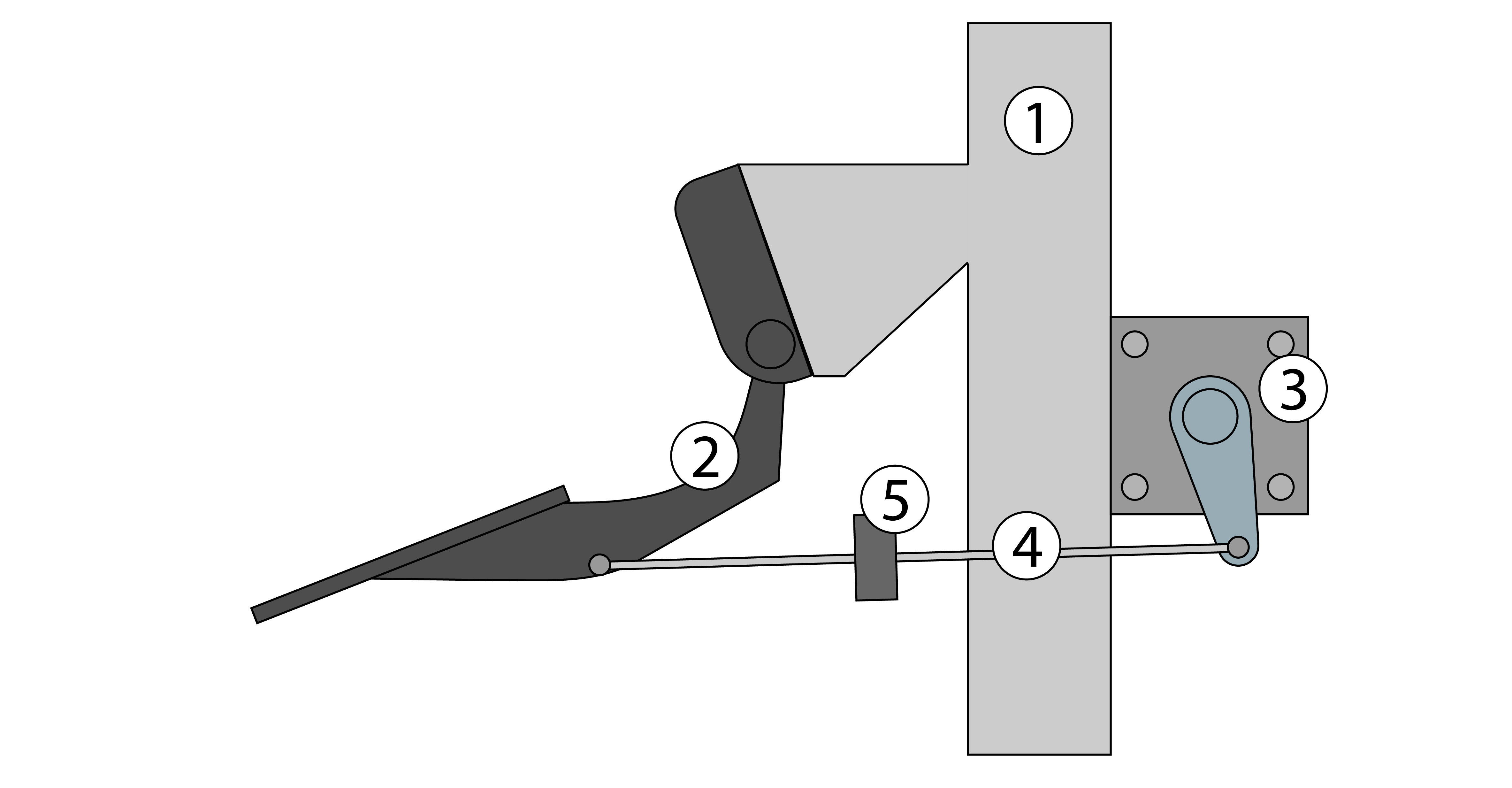}
    \caption{Schematic setup of the active accelerator pedal. On the frame (1), the pedal (2) is adjusted as well as the control loader (3), which is connected to the pedal via a rod (4) with a force cell (5) on it}
    \label{fig:pedals}
\end{figure}

\subsection{Hardware Design}
\begin{figure}
    \centering
    \includegraphics[width=0.9\linewidth]{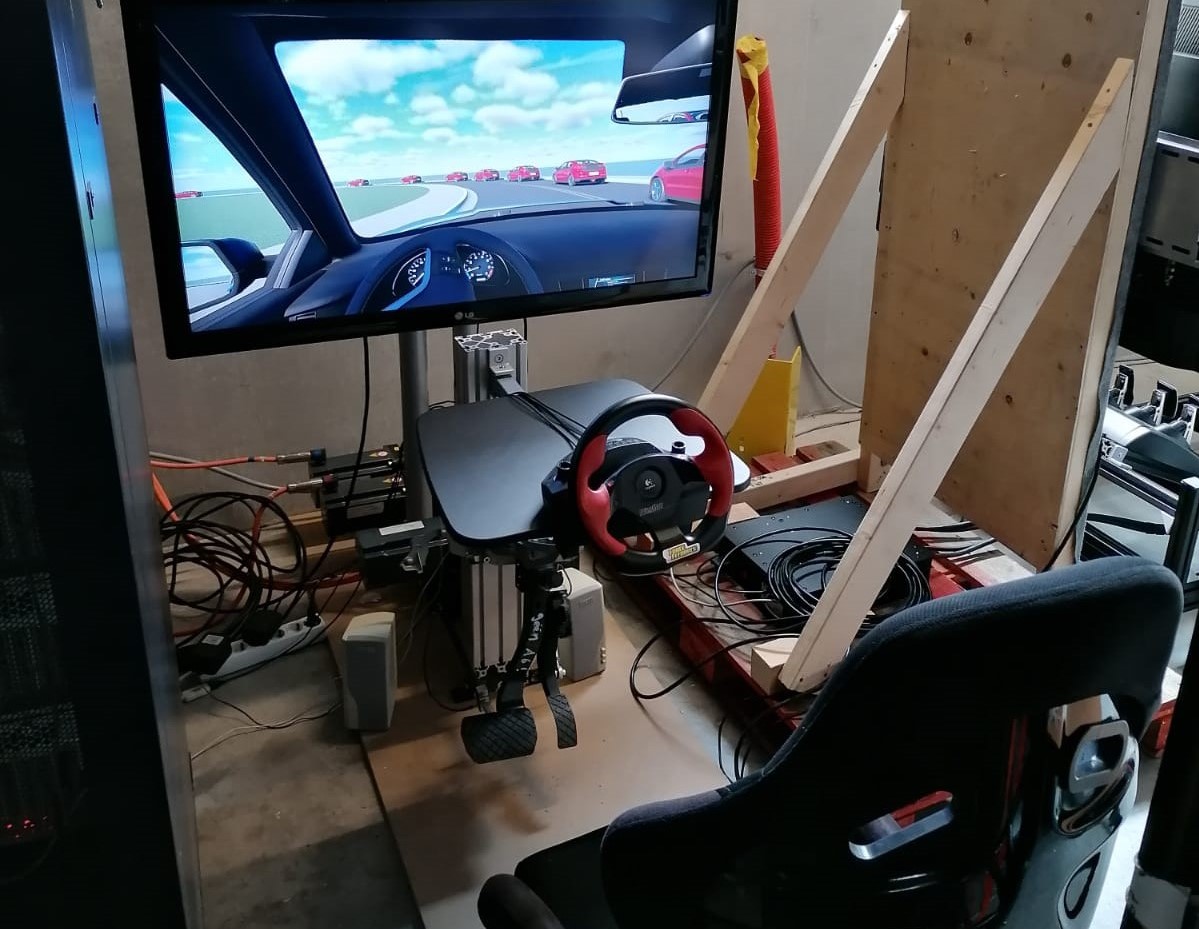}
    \caption{The fixed-base driving simulator setup.}
    \label{fig:sim}
\end{figure}

Figure \ref{fig:pedals} shows a picture of the active pedals which are made from Audi pedals. The pedals are connected to a servo motor via a metal rod, with an axial force sensor placed in the middle of it. This setup is called a control loading system. A feedback loop, closed with the axial force sensor, controls the force that the servo motor applies to the linkage. The pedal has the spring-damper behaviour, meaning that the pedal force is calculated as follows:

\begin{equation}
   F^\textrm{pedal}=K \times S^p+v^p \times b,
\end{equation}

where $K, S^p,v^p and b$ are the pedal stiffness, pedal position, pedal speed and pedal damping respectively. The active pedal setup was installed in a custom-built fixed-base simulator (Figure \ref{fig:sim}). The simulator consists of a car seat and a frame carrying the pedals, a steering wheel, a 55 inch monitor, speakers, and a set of computers on which the simulation runs. We used the IPG CarMaker software suite which was integrated with the commercially available Cruden Panthera software~\cite{Panthera2022}. This software setup combines the steering wheel and pedal control inputs, the audiovisual rendering, and the simulated scenario into one coherent real-time simulation. 

\section{Experiment Setup}
\subsection*{Scenario}
\begin{figure}
\centering
\includegraphics[width=0.95\linewidth]{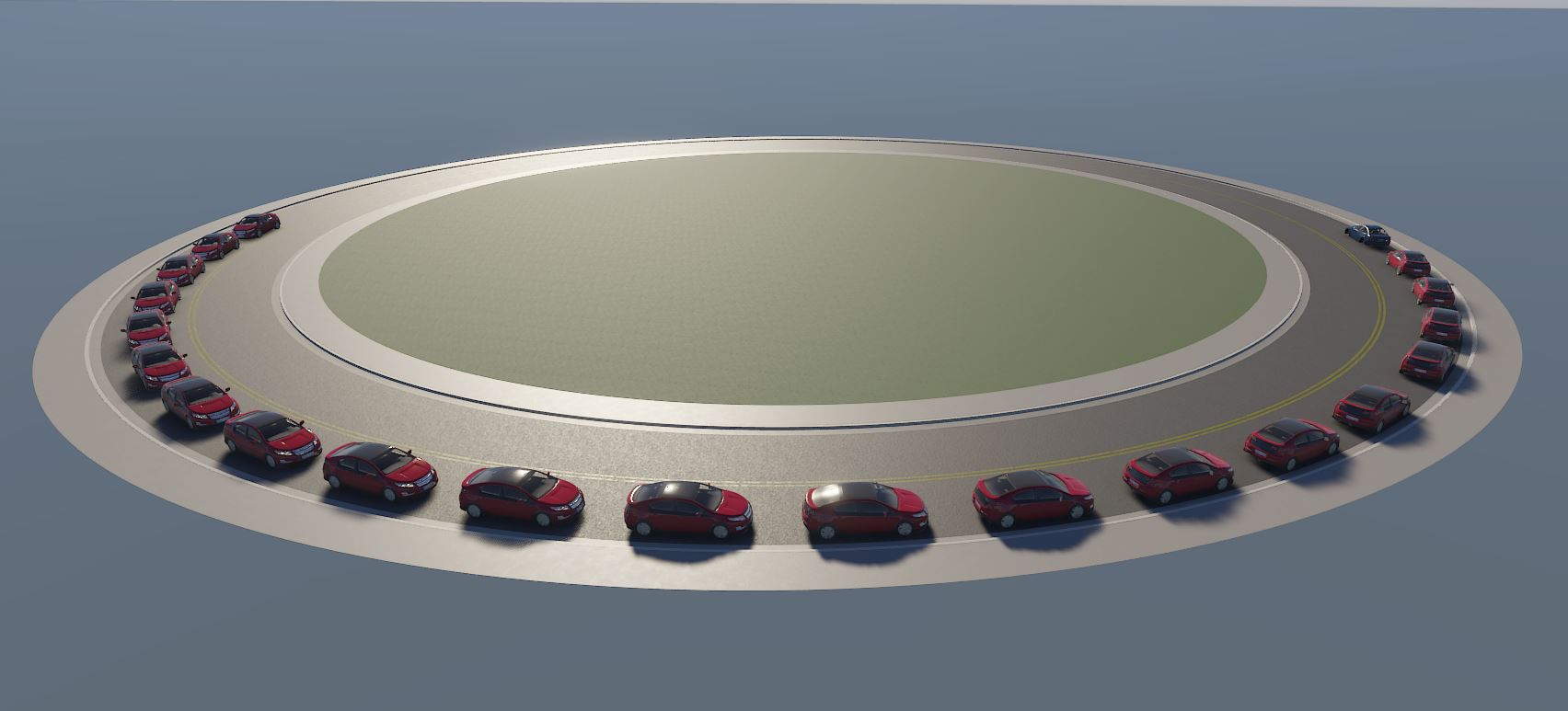}
\caption{Bird's-eye view of the simulated ring road environment. The blue car is the ego vehicle and the red cars are the simulated traffic cars.}
\label{fig:birdseye}
\end{figure}

We evaluated the haptic shared controller in a ring road scenario with a radius of 42 meters and 21 cars on it, including the ego vehicle (Figure~\ref{fig:birdseye}). This scenario is based on earlier studies that observe traffic jam formation on ring roads~\cite{sugiyama2008traffic, tadaki2013phase, stern2018dissipation}. We took the circumference and number of cars in our scenario from \textcite{stern2018dissipation} because the $\Delta x^0_k$ and $d_k$ parameters from equation \ref{eq:vcmd} are tuned specifically for this scenario. After pilot studies, we decided not to let the cars in the simulation start equally spaced around the ring road as in~\cite{sugiyama2008traffic, tadaki2013phase, stern2018dissipation}, but to concentrate them behind the ego vehicle at the start of the experiments. Because of this, there was an initial traffic jam that the ego vehicle drives into in the beginning of the experiment, allowing the traffic jam dissipation of the different means of control to be evaluated without waiting for the formation of a traffic jam. The 20 traffic cars are controlled by the IIDMACC driver model~\cite{treiber2013traffic}, with some minor adjustments provided by IPG CarMaker~\cite{IPG2020}.

\subsection*{Experimental conditions}
The experiments consisted of 3 driving sessions of 8 minutes during which we asked participants to drive on the simulated ring road. During each session, we evaluated participants' behavior in one of the three conditions:
\begin{itemize}
    \item \textit{Manual control}. In this condition, the participant fully controls the accelerator and brake pedal. 
    \item \textit{Haptic shared control}. In this condition, the participants' control input was combined with the input of the controller proposed by Stern et al.~\cite{stern2018dissipation} in the way described in Section II.
    \item \textit{Fully automated control}. In this condition, the accelerator and brake pedal are completely controlled by the input controller from Stern et al.~\cite{stern2018dissipation}. The participant controls the steering wheel and monitors the car. The participant is able to intervene by stepping on the pedals.
\end{itemize}

Before the start of the experiment, participants drove in the manual condition for two minutes to get used to the controls and the simulator. The order in which the conditions were tested was randomized across participants to counterbalance potential learning effects. During the manual and haptic condition, participants were instructed to control the steering wheel and the pedals themselves. In the automated condition, they were instructed only to steer and watch the road, as both pedals were fully automated. In this condition, participants were instructed to only interfere with the pedals when a situation was deemed unsafe. 

\subsection*{Silent automation failure}
In the haptic and fully automated conditions, a silent automation failure occurred after 8 minutes of driving. It simulated a real-life situation in which the camera system fails to detect the leading vehicle. We simulated this by sending a value of 1000 meters for the bumper-to-bumper gap to the velocity controller, resulting in a $v^{\textrm{cmd}}$ equal to $U$, causing the haptic accelerator pedal to decrease its stiffness and the automated pedal to get depressed until $U$ was reached. When this happened, the participant needed to intervene to prevent collision with the leading vehicle. After the participant regained control of the ego vehicle, we terminated the driving session.

\subsection*{Hypotheses}
We hypothesised that, compared to manual control, the haptic shared controller would increase the stability of the individual vehicle and the traffic flow as a whole, and reduce the traffic jam lifetime and the number of times the traffic jam is not dissipated. We also hypothesized that the haptic shared controller will result in higher throughput, compared to the manual driving condition. At the same time, we hypothesised that the fully automated controller will improve these metrics even further compared to the haptic shared controller. Finally, we hypothesised that the haptic shared controller will be safer than the fully automated controller in the event of silent automation failure, resulting in larger minimal gap after the failure and reduced number of collisions.

\subsection*{Statistical analyses}
To compare the three different means of control, for each participant we calculated metrics from the signals recorded during each session. These metrics quantified different aspects of the ego vehicle motion and the behavior of the entire vehicle platoon. To evaluate differences between the metrics that represent continuous variables, we performed paired t-tests. To compare discrete-valued metrics representing the occurrence of an event, we performed the McNemar test~\cite{mcnemar1947note}. For the t-test as well as the McNemar test, we deem the difference in metric values to be significant when the p-value is lower than 0.05.

\begin{figure*}
     \centering
     \begin{subfigure}[t!]{0.3\textwidth}
         \centering
         \includegraphics[width=\textwidth]{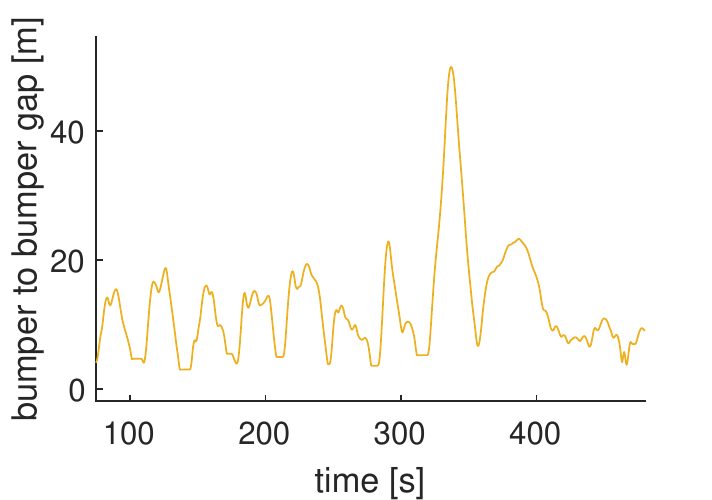}
         \caption{Manual}
         \label{fig:gapman}
     \end{subfigure}
     \hfill
     \begin{subfigure}[t!]{0.3\textwidth}
         \centering
         \includegraphics[width=\textwidth]{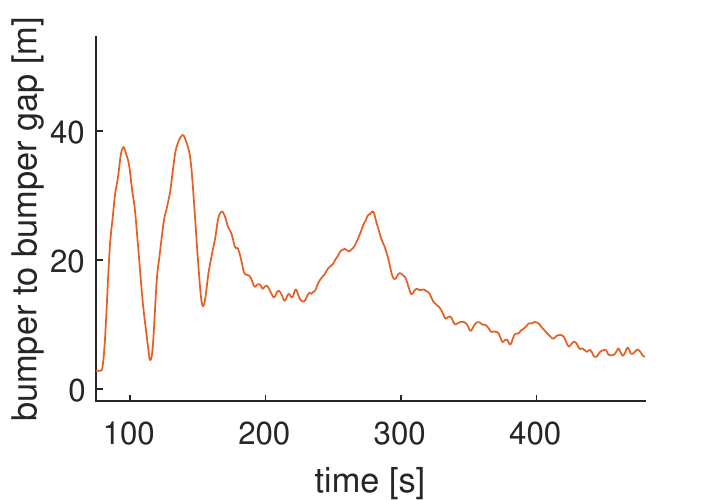}
         \caption{Haptic}
         \label{fig:gaphaptic}
     \end{subfigure}
     \hfill
     \begin{subfigure}[t!]{0.3\textwidth}
         \centering
         \includegraphics[width=\textwidth]{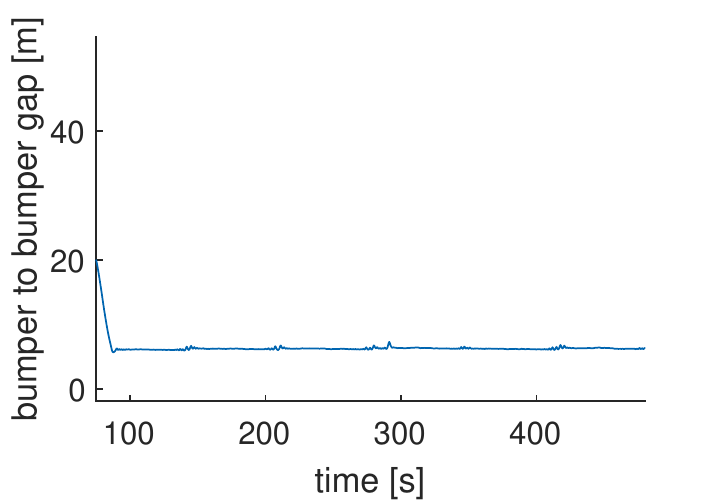}
         \caption{Automatic}
         \label{fig:gapautom}
     \end{subfigure}
        \caption{Bumper-to-bumper gap between the ego vehicle and the leading vehicle over time for a representative participant for each condition.}
        \label{fig:gapr26}
\end{figure*}

\begin{figure*}
     \centering
     \begin{subfigure}[t!]{0.3\textwidth}
         \centering
         \includegraphics[width=\textwidth]{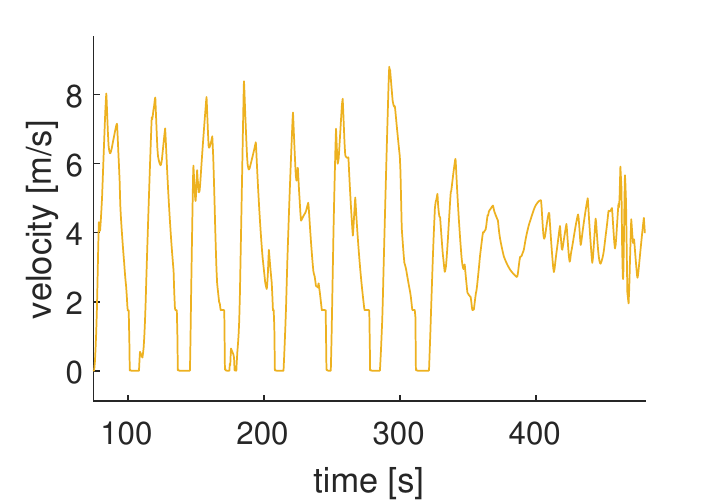}
         \caption{Manual}
         \label{fig:velman}
     \end{subfigure}
     \hfill
     \begin{subfigure}[t!]{0.3\textwidth}
         \centering
         \includegraphics[width=\textwidth]{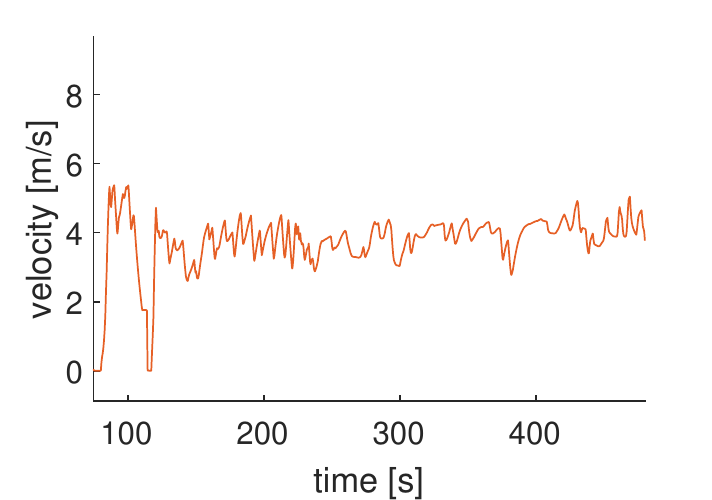}
         \caption{Haptic}
         \label{fig:velhaptic}
     \end{subfigure}
     \hfill
     \begin{subfigure}[t!]{0.3\textwidth}
         \centering
         \includegraphics[width=\textwidth]{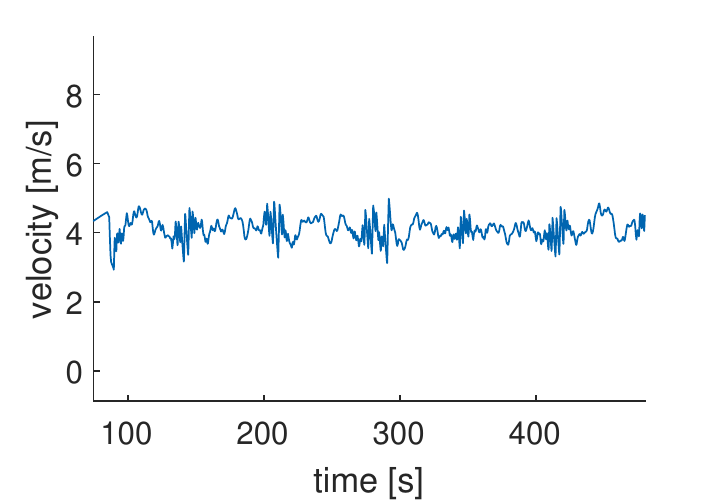}
         \caption{Automated}
         \label{fig:velautom}
     \end{subfigure}
        \caption{Velocity of the ego vehicle over time for a representative participant for each condition.}
        \label{fig:velr26}
\end{figure*}

\begin{figure*}
     \centering
     \begin{subfigure}[t!]{0.3\textwidth}
         \centering
         \includegraphics[width=\textwidth]{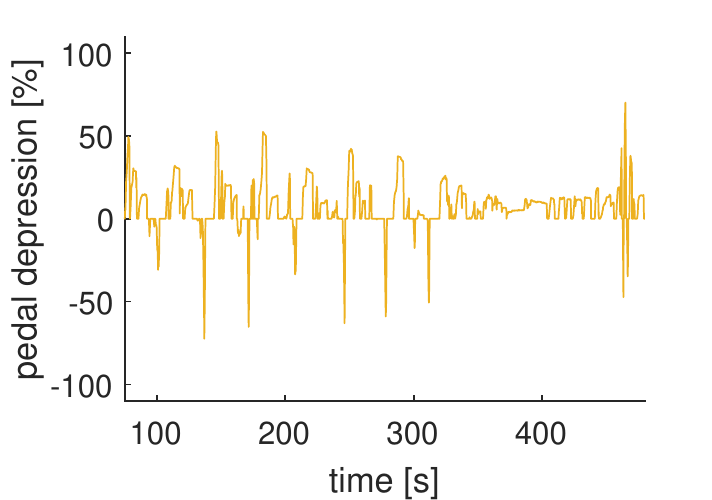}
         \caption{Manual}
         \label{fig:inputman}
     \end{subfigure}
     \hfill
     \begin{subfigure}[t!]{0.3\textwidth}
         \centering
         \includegraphics[width=\textwidth]{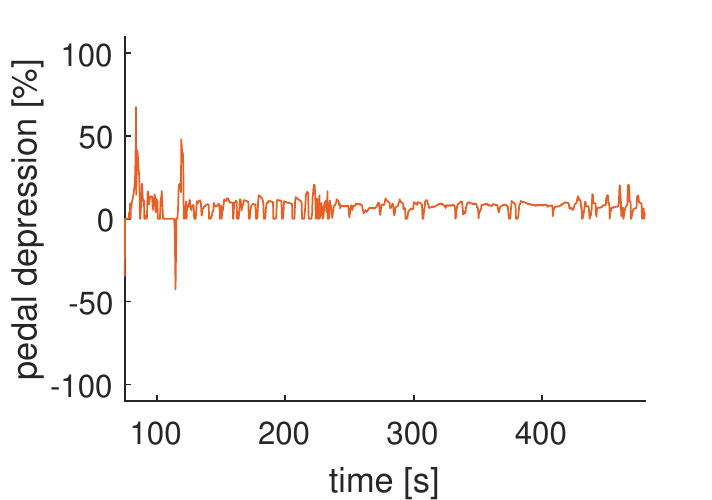}
         \caption{Haptic}
         \label{fig:inputhaptic}
     \end{subfigure}
     \hfill
     \begin{subfigure}[t!]{0.3\textwidth}
         \centering
         \includegraphics[width=\textwidth]{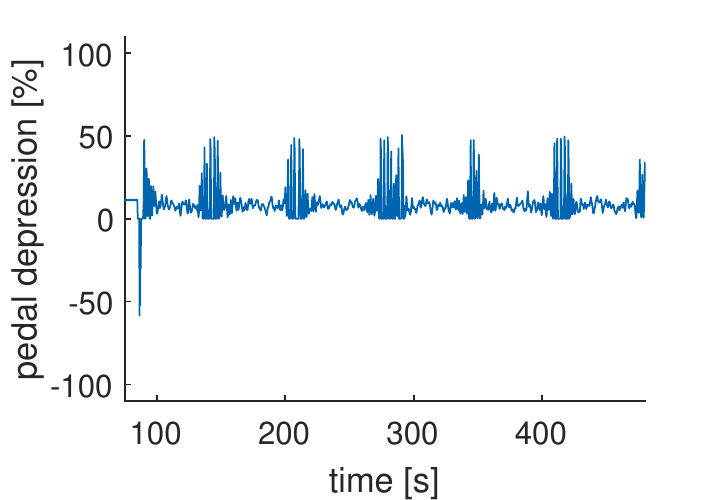}
         \caption{Automated}
         \label{fig:inputautom}
     \end{subfigure}
        \caption{Pedal depression of the ego vehicle over time for a representative participant for each condition. Positive values correspond to accelerator pedal depression, negative values correspond to brake pedal depression.}
        \label{fig:inputr26}
\end{figure*}

\begin{figure*}
     \centering
     \begin{subfigure}[t!]{0.3\textwidth}
         \centering
         \includegraphics[width=\textwidth]{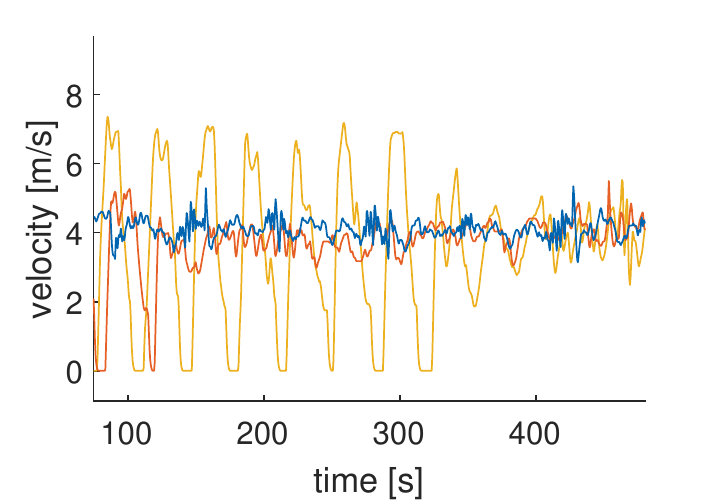}
         \caption{Following}
         \label{fig:intvelfol}
     \end{subfigure}
     \hfill
     \begin{subfigure}[t!]{0.3\textwidth}
         \centering
         \includegraphics[width=\textwidth]{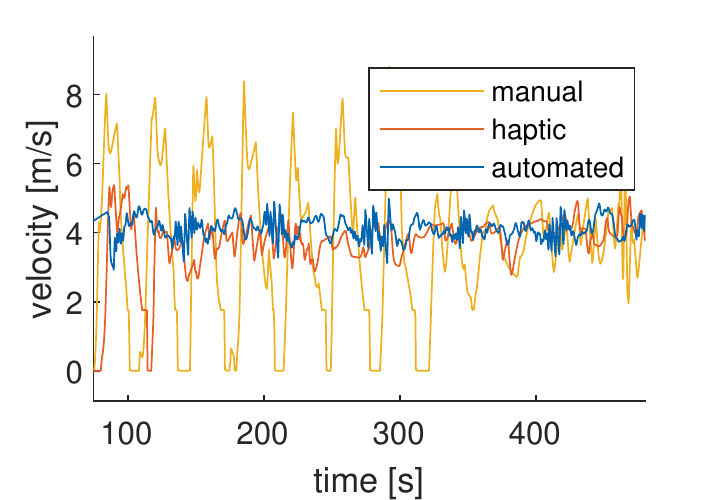}
         \caption{Ego}
         \label{fig:intvelego}
     \end{subfigure}
     \hfill
     \begin{subfigure}[t!]{0.3\textwidth}
         \centering
         \includegraphics[width=\textwidth]{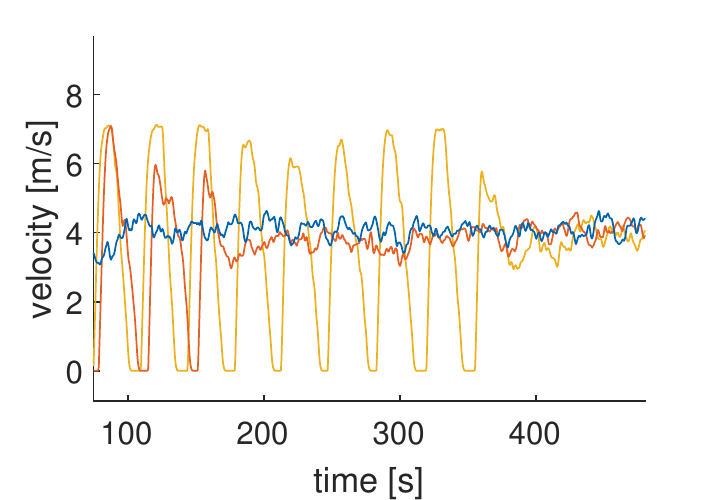}
         \caption{Leading}
         \label{fig:intvellead}
     \end{subfigure}
        \caption{Velocity of the following, ego, and leading vehicles for a representative participant for each condition.}
        \label{fig:intvelr26}
\end{figure*}

\section{Evaluation Results}
\subsection{Typical participant behavior}
Figures \ref{fig:gapr26}, \ref{fig:velr26}, \ref{fig:inputr26} and \ref{fig:intvelr26} illustrate behavior of a representative participant in the three conditions, starting after the transient period of 75 seconds\footnote{Figures illustrating behavior of all other participants can be found in the online supplementary material (\url{https://osf.io/rpwfx/?view_only=cb69f2295a2046a7bcf2565db97ce97c)}}. 

The dynamics of the gap between the ego vehicle and the leading vehicle (Figure \ref{fig:gapr26}) shows that in the manual and haptic cases, the driver is free to determine this gap, while in the automated case, the algorithm keeps the gap at a constant value of 6.5 meters. 

Velocity traces (Figure~\ref{fig:velr26}) further highlight qualitative differences between the three conditions. In the manual case (Figure~\ref{fig:velman}, velocity initially oscillates between standstill (0 m/s) and the speed limit (7 m/s). After approximately 320 seconds, the stop-and-go wave is dissipated and the car drives at a velocity that oscillates around $4m/s$. The haptic shared control condition (Figure~\ref{fig:velhaptic}) shows similar oscillatory behaviour at the start of the experiment, but this already stops after 120 seconds. Finally, the automatic condition (Figure~\ref{fig:velautom}) shows a velocity signal that oscillates around $4m/s$ during the entire experiment, meaning that the traffic jam that is present at the start of the experiment is dissipated in less than 75 seconds. 

The input signals of the ego vehicle (Figure \ref{fig:inputr26}) illustrate that (at least for this participant) the number of braking instances is reduced in the haptic shared control and fully automated conditions compared with the manual control condition. Interestingly, it also shows that the accelerator pedal positions for both the manual case and the automatic case are more extreme than those of the haptic case. For the manual and automatic case, the pedal depression often reaches 50\%, even after the traffic jam has already been dissipated. 

To illustrate how the behaviour of the ego vehicle influences the behaviour of the neighboring vehicles, we plotted their velocities next to each other (Figure \ref{fig:intvelr26}). The velocity profiles of the following (Figure~\ref{fig:intvelfol}) and leading (Figure~\ref{fig:intvellead}) vehicles are similar to those of the ego vehicle while their driving algorithms remain the same over the different conditions.

\subsection{Ego vehicle stability}
We analyzed stability of the ego vehicle in the three conditions, as quantified by the standard deviation of the ego vehicle velocity and the number of braking instances over each session.

\begin{figure}
    \centering
    \includegraphics[width=\linewidth]{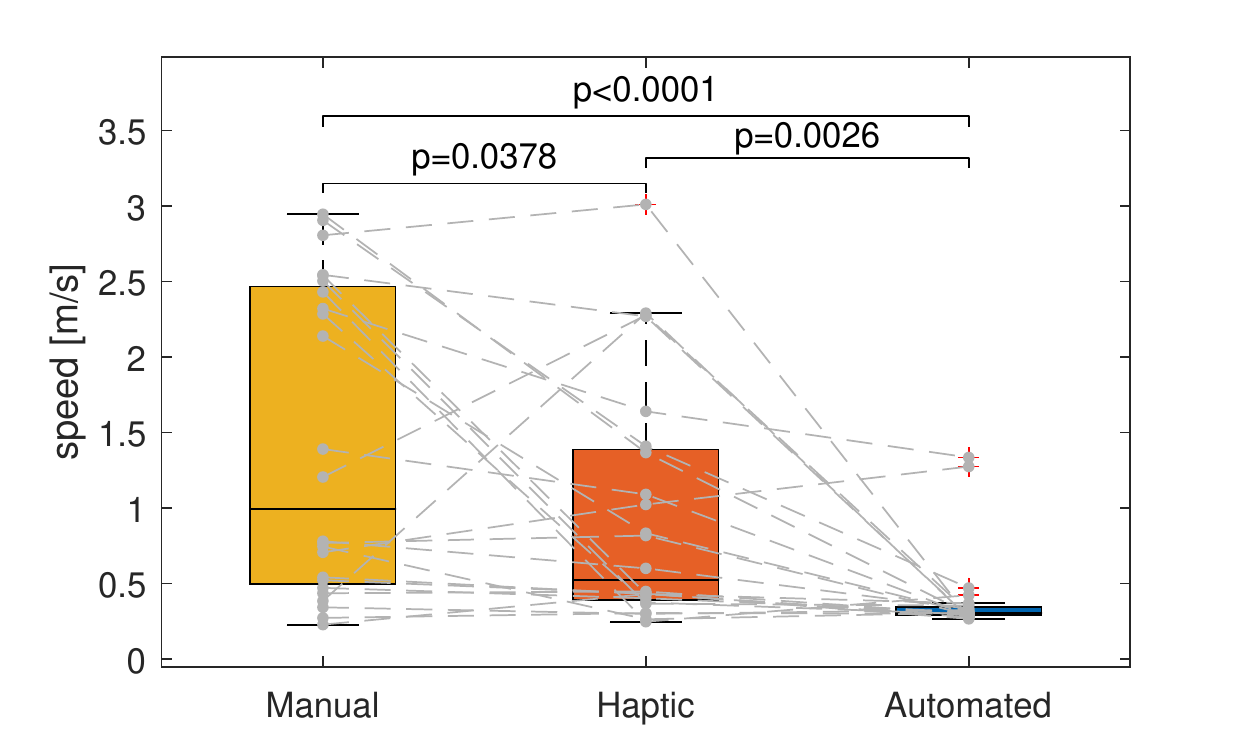}
    \caption{Standard deviation of the speed of the ego vehicle for the three different conditions}
    \label{fig:stdspeedego}
\end{figure}

Across all participants, the standard deviation of the ego vehicle speed was on average significantly smaller in the haptic shared control condition compared to the manual control condition ($p=0.0379, t=2.2028$, Figure~\ref{fig:stdspeedego}). The fully automated condition showed a further decrease ($p=0.0001, t=4.8916 $ and $p=0.00260, t=3.3727$ when comparing with the manual and haptic case respectively). 

\begin{figure}
    \centering
    \includegraphics[width=\linewidth]{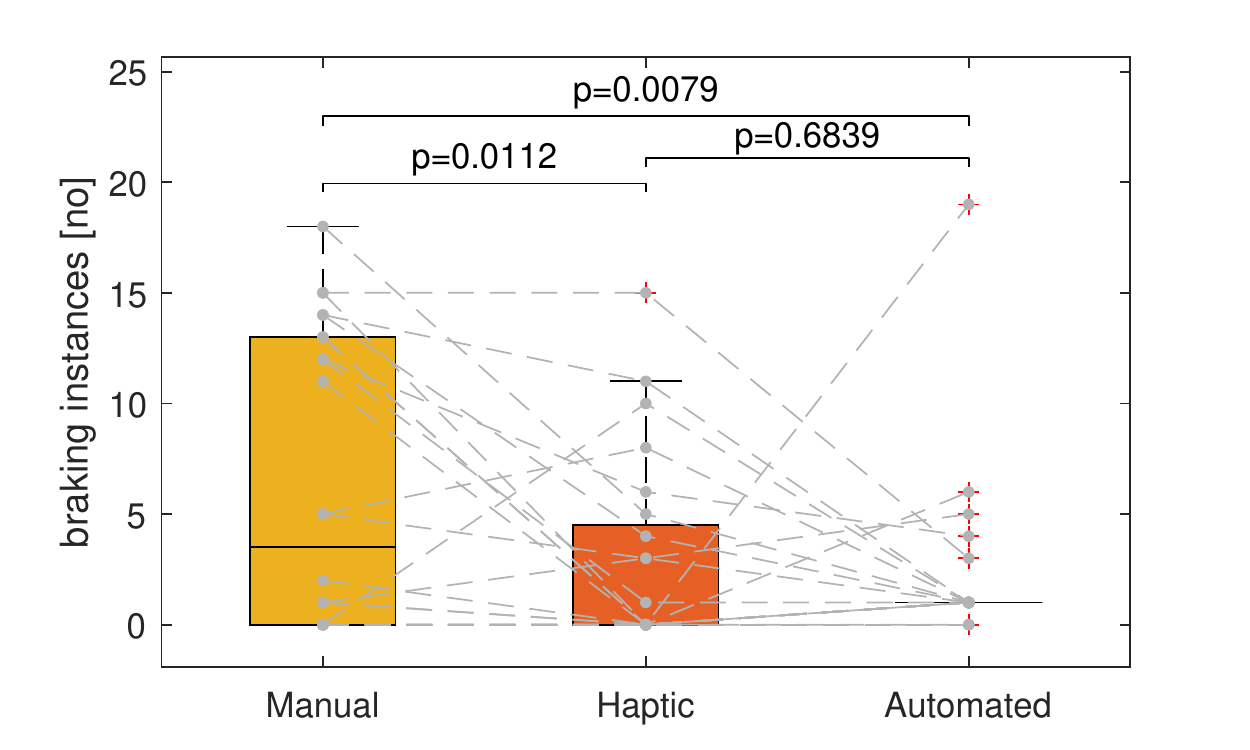}
    \caption{Number of braking instances of the ego vehicle for each condition.}
    \label{fig:braking}
\end{figure}

The number of braking instances of the ego vehicle in the haptic and automated conditions was significantly smaller than in the manual condition ($p=0.0112, t=-2.7550$, and $p=0.0079, t=-2.9061$, respectively; see also Figure~\ref{fig:braking}). However, there was no evidence that full automation reduces the number of braking instances compared to haptic shared control ($p= 0.6839, t=0.4123$).

\subsection{Platoon stabilization and traffic jam dissipation}
Stability of all the vehicles in the platoon was quantified by the average standard deviation of the velocity across all vehicles. Higher values of this metric would indicate larger velocity fluctuations across the platoon. 

\begin{figure}
    \centering
    \includegraphics[width=\linewidth]{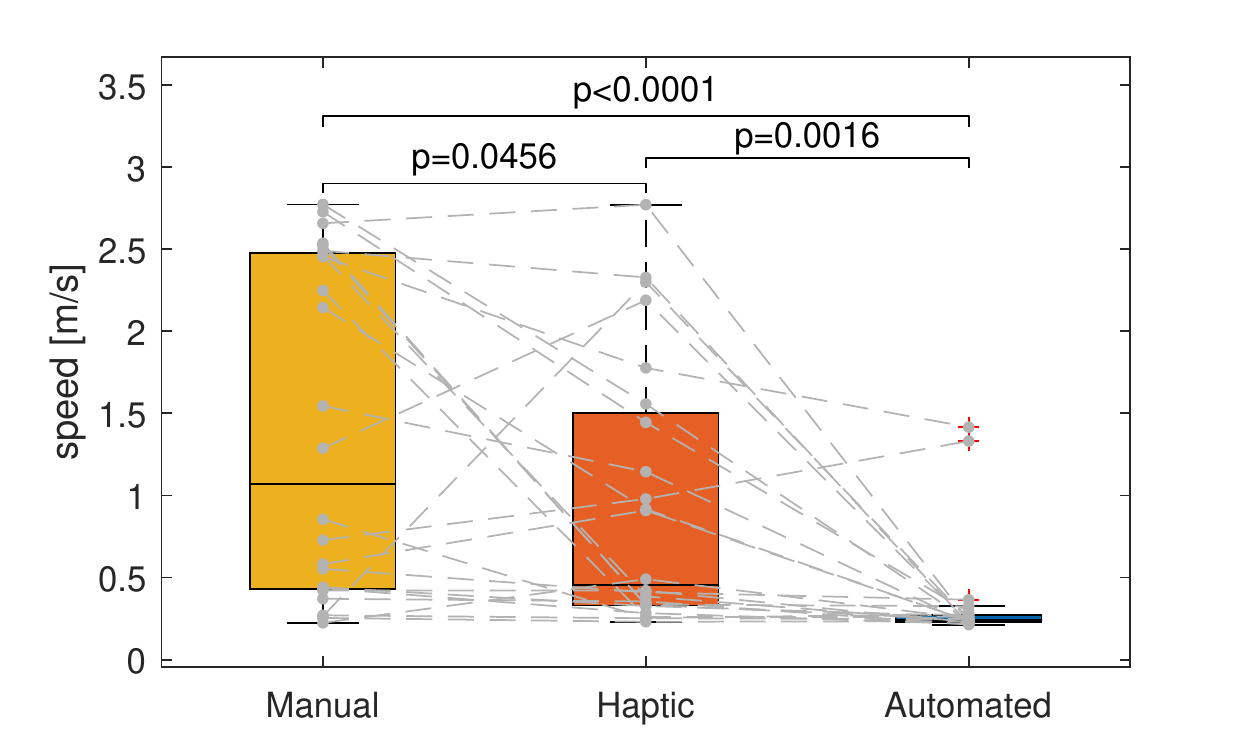}
    \caption{Standard deviation of the speed of the vehicles for the three different conditions}
    \label{fig:stdspeed}
\end{figure}

Haptic shared control reduced the standard deviation of the speed for the entire platoon significantly compared to the manual condition ($p=0.0456, t=2.1128$, Figure~\ref{fig:stdspeed}). Automation reduced it even further compared to both the haptic shared control as well as the manual case ($p=0.0016, t=3.5731$ and $p<0.0001, t=4.9012$ respectively).

Efficiency of the traffic jam dissipation was measured by traffic jam lifetime and the road throughput. We defined the lifetime of the traffic jam as the first moment (after the transient period of 75 seconds) at which none of the cars had zero speed anymore. For the start time of the traffic jam we took 75 seconds, as that was the moment the last vehicle took off. From this moment on, all the cars move and the initially imposed traffic jam is dissipated. If by the end of the session (480 seconds) the traffic jam was not dissolved, we considered its lifetime to be 405 seconds. 

\begin{figure}
    \centering
    \includegraphics[width=\linewidth]{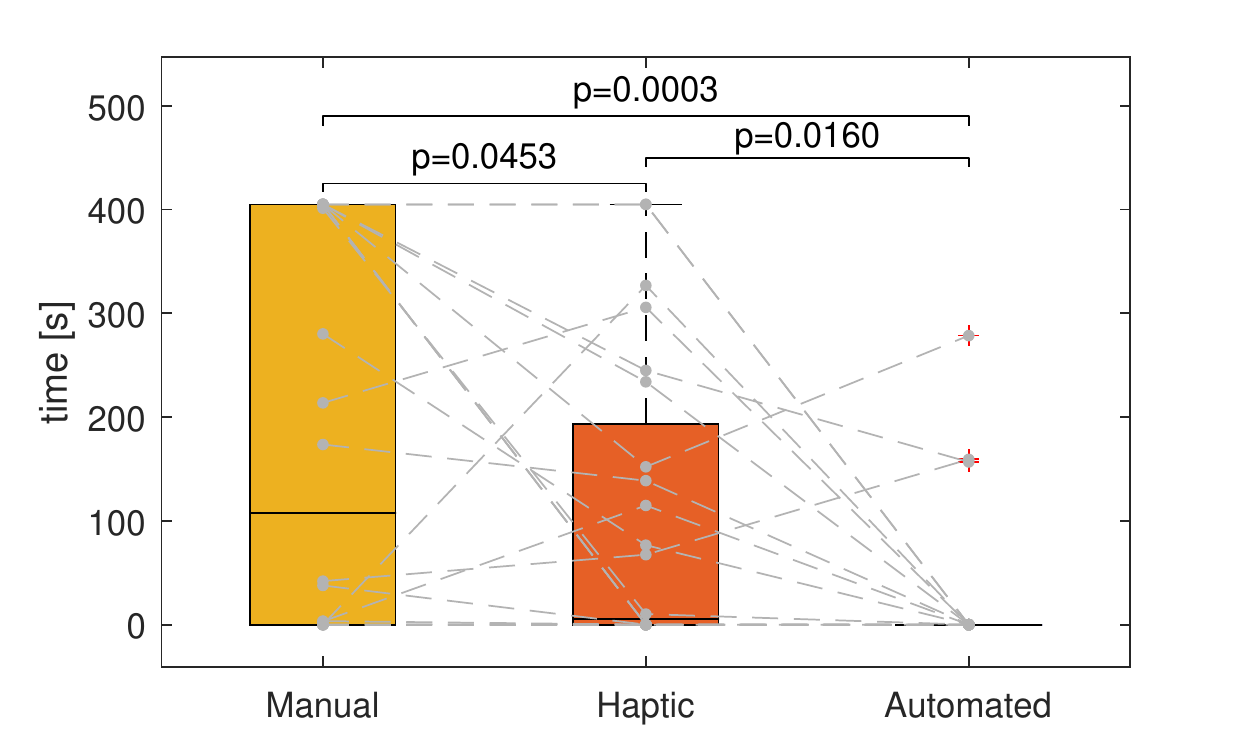}
    \caption{Lifetime of the initially imposed traffic jam for the three conditions}
    \label{fig:life}
\end{figure}

Lifetime of the initial traffic jam was lower in the haptic case compared to the manual case ($p=0.0453, t=2.116$, Figure \ref{fig:life}). In the automated case, the traffic jam lifetime was lower than in the manual case and the haptic shared control case ($p=0.0003, t=4.1849$ and $p=0.0160, t=2.5985$ respectively). These comparisons however are subject to the ceiling effect due to the limited duration of the experiment. To provide further insight into differences between conditions, we analyzed the number of participants which did not dissipate traffic jam by the end of the trial. In the manual control condition, 9 out of 24 participants did not manage to dissipate the traffic jam. In the haptic shared control case, it was not dissipated only by 2 out of 24 participants. In the automatic case, the jam was always dissipated. The differences between conditions are statistically significant when comparing haptic shared control to manual control ($\textrm{McNemar}\,\chi^2 =5.143, p=0.02334$), and full automation with manual control ($\textrm{McNemar}\,\chi^2 =7.111, p=0.007661$). There is no evidence of a significant difference when comparing haptic shared control with automation ($\textrm{McNemar}\,\chi ^2=0.5, p=0.4795$). 

We calculated road throughput by counting the number of cars that have passed the origin of the ring road during the driving session. This number was divided by 8 to obtain the average number of vehicles that drive past the start of the road per minute.

\begin{figure}
    \centering
    \includegraphics[width=1.08\linewidth]{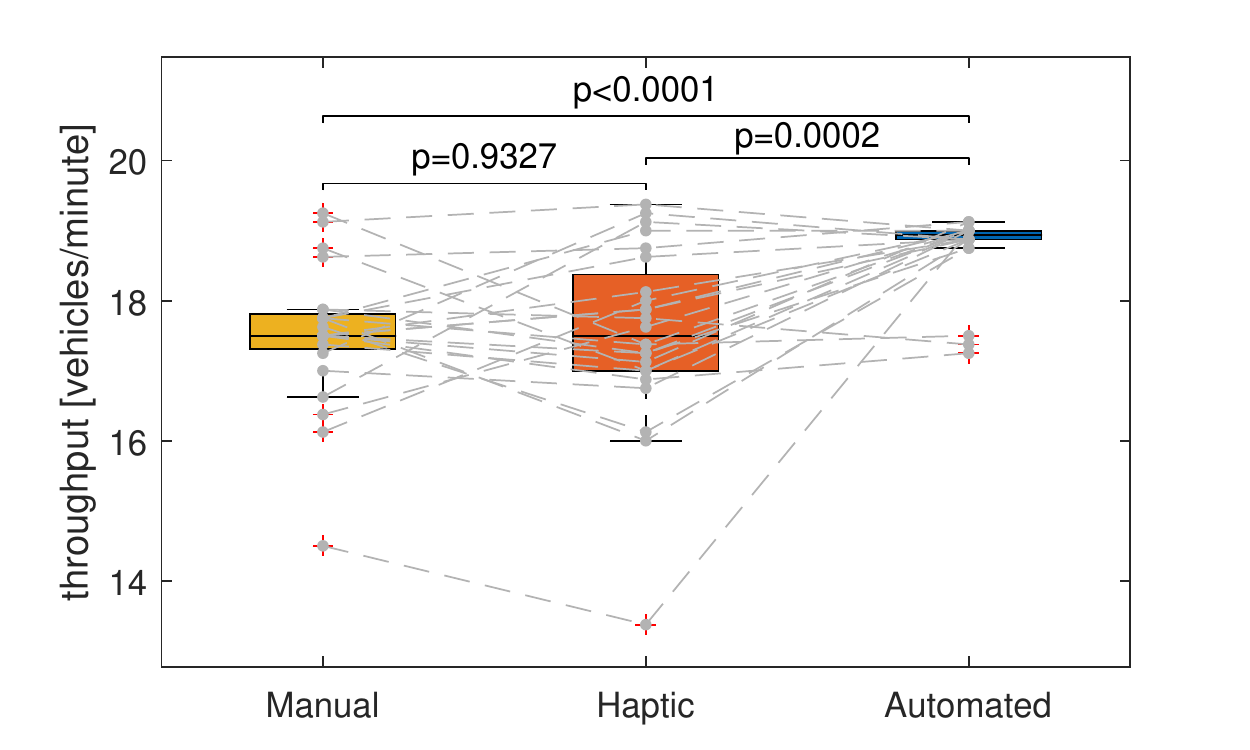}
    \caption{Throughput of vehicles during the three different conditions}
    \label{fig:throughput}
\end{figure}

The road throughput for the manual and haptic shared control case were in the same range, both in terms of median and variance (Figure~\ref{fig:throughput}). There was no evidence of a difference between these conditions ($p=0.9327, t=0.0853$). The throughput for the fully automated condition however was higher than for the manual and haptic conditions ($p=0.0002, t=4.3711$ and $p<0.0001, t=4.8840$, respectively).

\subsection{Safety}
Our setup including a silent automation failure occurring at the end of each trial allowed us to investigate how it was handled by the participants in each condition. Safety during the automation failure was measured based on the bumper-to-bumper gap between the ego vehicle and the leading vehicle (only for the haptic and the automated condition). We used the minimal bumper-to-bumper gap after the silent automation failure as a metric, as well as the occurrence of a collision (which was detected if the gap value reached 0). 

\begin{figure}
    \includegraphics[width=\linewidth]{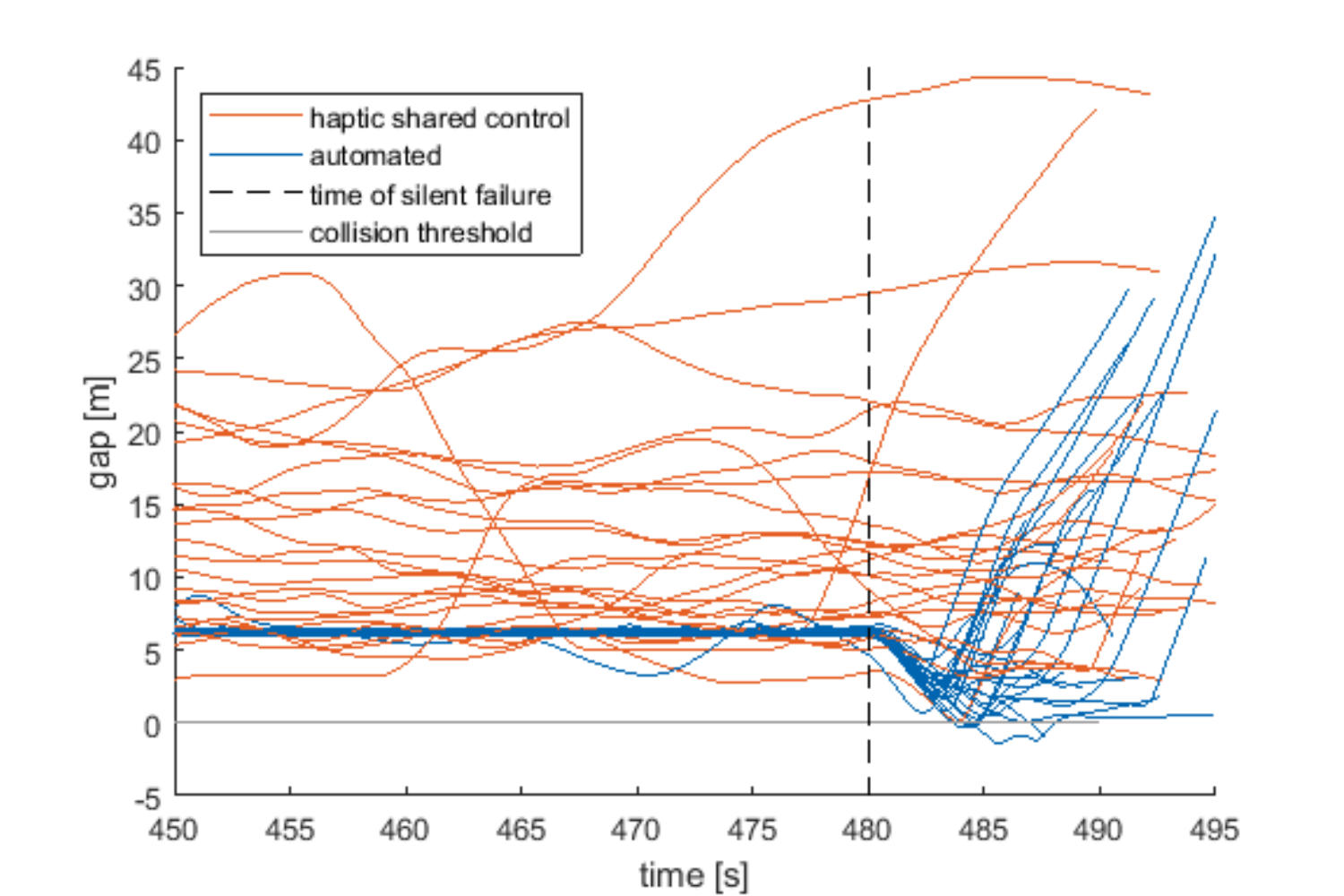}
    \captionof{figure}{Bumper to bumper gap between the ego vehicle and the leading vehicle from 450s until the end of the run for the haptic shared control and automated conditions}
    \label{fig:collision}
\end{figure}

In the fully automated condition, the simulated failure in the leading vehicle tracking meant that the automation depressed the accelerator pedal. This led to decreasing bumper-to-bumper gap between the leading vehicle and the ego vehicle right after the failure (Figure \ref{fig:collision}). To avoid a collision, the driver needed to intervene and press the brake, after which the gap increased again. In 5 out of 24 participants, this automation failure caused a collision. 

\begin{figure}
    \centering
    \includegraphics[width=\linewidth]{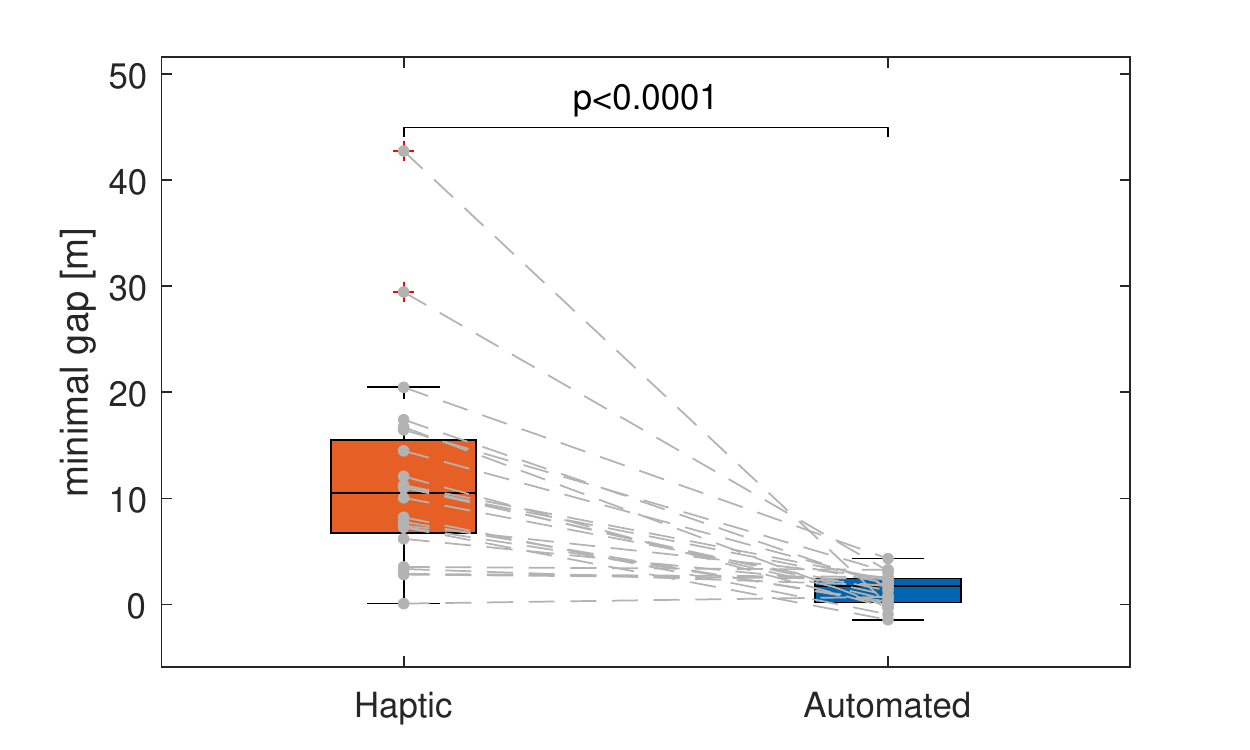}
    \captionof{figure}{Minimal bumper-to-bumper gap between the ego vehicle and the leading vehicle for the haptic and automated condition after the silent automation failure}
    \label{fig:minimalgap}
\end{figure}

For the haptic controller, in some of the cases the gap reduces after the silent failure, but this reduction was never as drastic as in the automated case and values close to 0 were never reached (Figure \ref{fig:collision}). This is further illustrated by the minimal value the gap reached in 15 seconds after the failure (\ref{fig:minimalgap}). This minimal gap value was significantly larger in the haptic condition compared to the automated condition ($p<0.0001, t=5.3917$). However, there was no evidence of a decrease in the number of collisions in the haptic condition compared to the automated condition ($\textrm{McNemar}\,\chi^2=3.2, p=0.073638$).

\begin{figure}
    \centering
    \includegraphics[width=\linewidth]{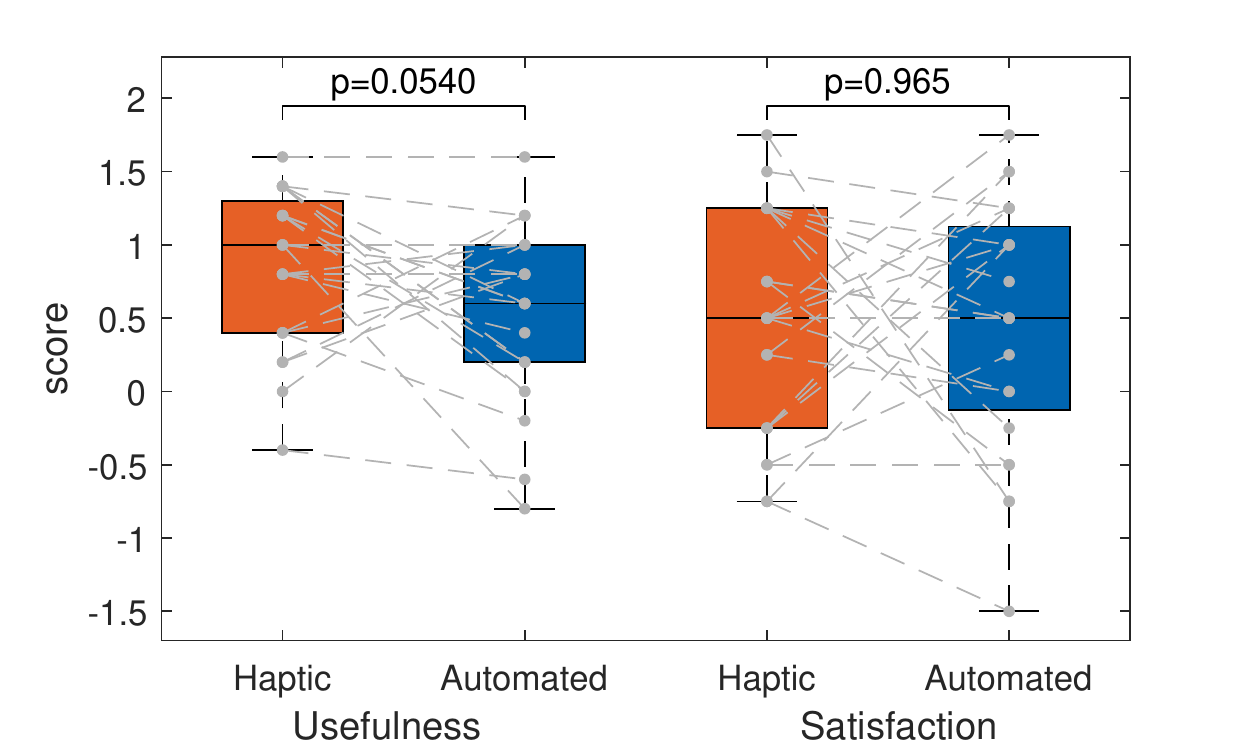}
    \captionof{figure}{Acceptance scores for the haptic shared control system and the automated system}
    \label{fig:acceptance}
\end{figure}

\subsection{User acceptance}
Subjective user acceptance of the haptic and fully automated controllers was measured using the Van der Laan questionnaire~\cite{van1997simple}, which quantifies how the participants perceive a system in terms of usefulness and satisfaction. The questionnaire consists of 9 questions where the participants grade the system on a five-point scale from -2 to 2. 

Both haptic shared control and full automation were positively evaluated by the participants (Figure~\ref{fig:acceptance}). Specifically, the means of the satisfaction and usefulness scores for both systems are significantly larger than zero, which means that the systems were rated as satisfactory (haptic $p=0.0041, t=3.1855$, fully automated $p=0.0125, t=2.7080$) and useful (haptic $p<0.0001, t=8.0656$, fully automated $p=0.0001, t=4.5859$). We found no evidence of a difference in subjective usefulness ($p=0.0541, t=2.030$) and user satisfaction ($p=0.9650, t=0.044$) between the haptic shared control system and the fully automated system.

\section{Discussion}
Our results showed that haptic shared control stabilised traffic and dissipated stop-and-go waves faster and more often than manual control. However, we found no evidence that haptic shared control resulted in improved road throughput. Full automation improved vehicle stability and traffic jam dissipation as well as vehicle throughput, compared to both manual control and haptic shared control. However, the fully automated system was less safe than haptic shared control in the event of the silent automation failure, as measured by the minimal gap to the leading vehicle. Both the haptic shared controller and the fully automated controller were rated positively by the users in terms of usefulness and satisfaction, and there was no evidence of a difference between the acceptance scores of the haptic and automatic systems. The fact that the achieved traffic jam dissipation performance of the haptic shared controller lies between the manual control case and the automated case is in line with our hypotheses. It is also in line with the essence of haptic shared control to blend manual control and automation.

The improved performance and stability that the automation achieves over the manual control case are in line with the findings by \textcite{stern2018dissipation}, which evaluated the same ring road scenario. Importantly, \textcite{stern2018dissipation} managed to obtain much higher mean speeds (about 7.5 m/s) than our study (not higher than 4 m/s). This difference can be attributed to the fact that \textcite{stern2018dissipation} instructed the human drivers to keep a much smaller gap than they would normally do. We did not program the simulated human drivers in this study to keep a tight gap, and neither did we instruct the human participants to do so because this is not in line with the driving behaviour of regular human drivers. The comparatively low driving speeds in our experiment could have led to a ceiling effect for the mean speed and therefore throughput, which could be the reason why we did not observe differences in these metrics between the manual case and the haptic shared control case. 

Our findings are also in line with \textcite{jiang2021robust}, who evaluate a similar ring road scenario as \cite{stern2018dissipation}, but with a shared algorithm instead of a fully automated algorithm. Their study shows that this sharing algorithm is able to stabilize traffic just like the automated controller from \cite{stern2018dissipation}. However, \textcite{jiang2021robust} have not evaluated their controller in a human-in-the-loop experiment, only demonstrating it with simulated drivers. Their algorithm manages to stabilize traffic for every type of human driving behaviour that was evaluated. We did not find this: some traffic jams were not solved for the haptic case. This difference can be explained either by discrepancies between the simulated and real humans, or by the different controller type: we used a haptic shared controller, where the human can always fully overrule the automation. On the other hand, as mentioned in the introduction, the non-haptic shared control algorithm from \cite{jiang2021robust} is designed in a way that the human driver only controls half of the inputs. While this leads to increased stability, it could potentially lead to reduced acceptance when driver and automation actions are misaligned, or reduced safety margins or even accidents in case of automation failures.

Our finding that the silent automation failure can result in decreased bumper-to-bumper gap and collisions in the automated case is in line with the previous findings about vigilance decrement and increased reaction times~\cite{parasuraman1987human, rudin2004behavioural}. This result, together with the finding that haptic shared control improves safety compared to full automation, also resonates with the study by \textcite{flemisch2008cooperative} which evaluated a similar scenario for haptic shared control for lateral vehicle motion to illustrate the advantage of keeping the driver in the loop.

The main limitations of the current study concern controller tuning and translating the finding to more realistic environments. First, the controller that calculates the target pedal position in the fully automated condition was tuned to have high gains, with the purpose of efficient dissipation of the traffic jam. However, high gains in some cases resulted in oscillatory behaviour; even when there was no phantom traffic jam, the accelerator pedal moved relatively aggressively, as compared to the haptic and manual case (Figure~\ref{fig:inputautom}). This behaviour allowed the controller to keep the bumper-to-bumper gap close to the target value (Figure~\ref{fig:gapautom}) but is likely suboptimal in terms of perceived motion comfort and fuel consumption. We therefore recommend that future studies investigate how tuning the automation's controller gains can balance dissipating the traffic jams and minimizing acceleration/velocity oscillations.

Second, this study has only shown the efficacy of haptic shared control for a simple artificial ring road scenario at low speeds. Although this proof-of-concept scenario shows the promise of haptic shared control, it does not simply generalize to real-world scenarios in several ways. First, phantom traffic jams occur on highways where cars drive at speeds in the range of 80-120 km/h. In this scenario, velocities range from 0 to 20 km/h. The ring road itself is also not representative of actual highways, which are usually straight and multi-lane. Furthermore, in a multi-lane scenario, the operation of longitudinal driving automation could be affected by lateral driving automation (e.g. lane-keep assist). Finally, our implementation of the silent automation failure is much simplified. For these reasons, an important direction for future research is evaluating our controller in more realistic scenarios, studying the effects of higher speeds, lateral automation, and more realistic automation failures. 
To the best of our knowledge this research is the first to investigate the potential for haptic shared control to dissipate phantom traffic jams, but it should not be the last if we want to gain better insight into the effectiveness of this means of traffic control.

\section{Conclusion}
This paper proposes a haptic shared control system designed to dissipate phantom traffic jams. We evaluated the system against manual control and full automation in a human-in-the-loop driving simulator experiment. For the experimental conditions studied, we conclude that: 
\begin{itemize}
    \item Haptic shared control dissipates phantom traffic jams faster and more often compared to manual control by a human alone.
    \item Full automation dissipates traffic jam faster than haptic shared control but leads to increased occurrence  of unsafe situations in case of a silent automation failure. 
\end{itemize}

We conclude that haptic shared control shows promise for mitigating phantom traffic jams while preventing safety risks associated with full driving automation. 

\section*{Acknowledgment}

The authors would like to thank Cruden Driving Simulators B.V. for facilitating hardware, software, and knowledge, all of which were necessary for building the experimental setup used in this study.

\ifCLASSOPTIONcaptionsoff
  \newpage
\fi

\printbibliography[title=\centering \small {\normalfont{References}}]

@article{mulder2010design,
  title={Design of a haptic gas pedal for active car-following support},
  author={Mulder, Mark and Abbink, David A and Van Paassen, Marinus M and Mulder, Max},
  journal={IEEE Transactions on Intelligent Transportation Systems},
  volume={12},
  number={1},
  pages={268--279},
  year={2010},
  publisher={IEEE}
}

@article{stern2018dissipation,
  title={Dissipation of stop-and-go waves via control of autonomous vehicles: Field experiments},
  author={Stern, Raphael E and Cui, Shumo and Delle Monache, Maria Laura and Bhadani, Rahul and Bunting, Matt and Churchill, Miles and Hamilton, Nathaniel and Pohlmann, Hannah and Wu, Fangyu and Piccoli, Benedetto and others},
  journal={Transportation Research Part C: Emerging Technologies},
  volume={89},
  pages={205--221},
  year={2018},
  publisher={Elsevier}
}

@article{goldmann2020economic,
  title={Economic implications of phantom traffic jams: evidence from traffic experiments},
  author={Goldmann, Kathrin and Sieg, Gernot},
  journal={Transportation Letters},
  volume={12},
  number={6},
  pages={386--390},
  year={2020},
  publisher={Taylor \& Francis}
}

@article{mulder2008haptic,
  title={Haptic gas pedal feedback},
  author={Mulder, Max and Mulder, M and Van Paassen, MM and Abbink, DA},
  journal={Ergonomics},
  volume={51},
  number={11},
  pages={1710--1720},
  year={2008},
  publisher={Taylor \& Francis}
}

@incollection{bainbridge1983ironies,
  title={Ironies of automation},
  author={Bainbridge, Lisanne},
  booktitle={Analysis, design and evaluation of man--machine systems},
  pages={129--135},
  year={1983},
  publisher={Elsevier}
}

@article{lee2019phantom,
  title={Phantom Traffic: Platoon Formed at Low Traffic Density},
  author={Lee, Jun and Kim, Jae Hun},
  journal={Journal of Transportation Engineering, Part A: Systems},
  volume={145},
  number={2},
  pages={04018082},
  year={2019},
  publisher={American Society of Civil Engineers}
}

@article{tadaki2013phase,
  title={Phase transition in traffic jam experiment on a circuit},
  author={Tadaki, Shin-ichi and Kikuchi, Macoto and Fukui, Minoru and Nakayama, Akihiro and Nishinari, Katsuhiro and Shibata, Akihiro and Sugiyama, Yuki and Yosida, Taturu and Yukawa, Satoshi},
  journal={New Journal of Physics},
  volume={15},
  number={10},
  pages={103034},
  year={2013},
  publisher={IOP Publishing}
}

@inproceedings{jamson2013design,
  title={The design of haptic gas pedal feedback to support eco-driving},
  author={Jamson, AH and Hibberd, Daryl L and Merat, Natasha},
  booktitle={Proceedings of the Seventh International Driving Symposium on Human Factors in Driver Assessment, Training, and Vehicle Design},
  pages={264--270},
  year={2013},
  organization={University of Iowa}
}

@article{wu2019tracking,
  title={Tracking vehicle trajectories and fuel rates in phantom traffic jams: Methodology and data},
  author={Wu, Fangyu and Stern, Raphael E and Cui, Shumo and Delle Monache, Maria Laura and Bhadani, Rahul and Bunting, Matt and Churchill, Miles and Hamilton, Nathaniel and Piccoli, Benedetto and Seibold, Benjamin and others},
  journal={Transportation Research Part C: Emerging Technologies},
  volume={99},
  pages={82--109},
  year={2019},
  publisher={Elsevier}
}

@article{treiber2013traffic,
  title={Traffic flow dynamics},
  author={Treiber, Martin and Kesting, Arne},
  journal={Traffic Flow Dynamics: Data, Models and Simulation, Springer-Verlag Berlin Heidelberg},
  year={2013},
  publisher={Springer}
}

@article{sugiyama2008traffic,
  title={Traffic jams without bottlenecks—experimental evidence for the physical mechanism of the formation of a jam},
  author={Sugiyama, Yuki and Fukui, Minoru and Kikuchi, Macoto and Hasebe, Katsuya and Nakayama, Akihiro and Nishinari, Katsuhiro and Tadaki, Shin-ichi and Yukawa, Satoshi},
  journal={New journal of physics},
  volume={10},
  number={3},
  pages={033001},
  year={2008},
  publisher={IOP Publishing}
}

@inproceedings{kreidieh2018dissipating,
  title={Dissipating stop-and-go waves in closed and open networks via deep reinforcement learning},
  author={Kreidieh, Abdul Rahman and Wu, Cathy and Bayen, Alexandre M},
  booktitle={2018 21st International Conference on Intelligent Transportation Systems (ITSC)},
  pages={1475--1480},
  year={2018},
  organization={IEEE}
}

@article{hoogendoorn2013assessment,
  title={Assessment of dynamic speed limits on freeway A20 near Rotterdam, Netherlands},
  author={Hoogendoorn, SP and Daamen, W and Hoogendoorn, RG and Goemans, JW},
  journal={Transportation research record},
  volume={2380},
  number={1},
  pages={61--71},
  year={2013},
  publisher={SAGE Publications Sage CA: Los Angeles, CA}
}

@article{parasuraman1987human, title={Human-computer monitoring}, author={Parasuraman, Raja}, journal={Human Factors}, volume={29}, number={6}, pages={695--706}, year={1987}, publisher={SAGE Publications Sage CA: Los Angeles, CA} }

@article{abbink2010neuromuscular,
  title={Neuromuscular analysis as a guideline in designing shared control},
  author={Abbink, David A and Mulder, Mark},
  journal={Advances in Haptics, Mehrdad Hosseini Zadeh (Ed.), ISBN: 9789533070933},
  year={2010},
  publisher={InTech}
}

@article{flemisch2008cooperative, title={Cooperative control and active interfaces for vehicle assitsance and automation}, author={Flemisch, Frank and Kelsch, Johann and L{\"o}per, Christan and Schieben, Anna and Schindler, Julian and Heesen, Matthias}, year={2008} }

@article{azzi2011eco,
  title={Eco-driving performance assessment with in-car visual and haptic feedback assistance},
  author={Azzi, Slim and Reymond, Gilles and M{\'e}rienne, Fr{\'e}d{\'e}ric and Kemeny, Andras},
  journal={Journal of Computing and Information Science in Engineering},
  volume={11},
  number={4},
  year={2011},
  publisher={American Society of Mechanical Engineers Digital Collection}
}

@article{adell2008auditory,
  title={Auditory and haptic systems for in-car speed management--A comparative real life study},
  author={Adell, Emeli and V{\'a}rhelyi, Andr{\'a}s and Hj{\"a}lmdahl, Magnus},
  journal={Transportation research part F: traffic psychology and behaviour},
  volume={11},
  number={6},
  pages={445--458},
  year={2008},
  publisher={Elsevier}
}

@article{mahmud2012possible,
  title={Possible causes \& solutions of traffic jam and their impact on the economy of Dhaka City},
  author={Mahmud, Khaled and Gope, Khonika and Chowdhury, Syed Mustafizur Rahman},
  journal={J. Mgmt. \& Sustainability},
  volume={2},
  pages={112},
  year={2012},
  publisher={HeinOnline}
}

@inproceedings{vcivcic2018traffic, title={Traffic regulation via individually controlled automated vehicles: a cell transmission model approach}, author={{\v{C}}i{\v{c}}i{\'c}, Mladen and Johansson, Karl Henrik}, booktitle={2018 21st International Conference on Intelligent Transportation Systems (ITSC)}, pages={766--771}, year={2018}, organization={IEEE} }

@article{gunter2020commercially, title={Are commercially implemented adaptive cruise control systems string stable?}, author={Gunter, George and Gloudemans, Derek and Stern, Raphael E and McQuade, Sean and Bhadani, Rahul and Bunting, Matt and Delle Monache, Maria Laura and Lysecky, Roman and Seibold, Benjamin and Sprinkle, Jonathan and others}, journal={IEEE Transactions on Intelligent Transportation Systems}, year={2020}, publisher={IEEE} }

@inproceedings{son2006solution,
  title={A solution for the dropout problem in adaptive cruise control range sensors},
  author={Son, Bongsoo and Kim, Taehyung and Shin, YongEun},
  booktitle={International Conference on Embedded and Ubiquitous Computing},
  pages={979--987},
  year={2006},
  organization={Springer}
}

@book{nilsson1996safety,
  title={Safety effects of adaptive cruise controls in critical traffic situations},
  author={Nilsson, Lena},
  year={1996},
  publisher={Statens v{\"a}g-och transportforskningsinstitut., VTI s{\"a}rtryck 265}
}

@article{rudin2004behavioural,
  title={Behavioural adaptation to adaptive cruise control (ACC): implications for preventive strategies},
  author={Rudin-Brown, Christina M and Parker, Heather A},
  journal={Transportation Research Part F: Traffic Psychology and Behaviour},
  volume={7},
  number={2},
  pages={59--76},
  year={2004},
  publisher={Elsevier}
}

@article{jiang2021robust,
  title={Robust traffic wave damping via shared control},
  author={Jiang, Jingjing and Astolfi, Alessandro and Parisini, Thomas},
  journal={Transportation Research Part C: Emerging Technologies},
  volume={128},
  pages={103110},
  year={2021},
  publisher={Elsevier}
}

@inproceedings{kim2015demonstration,
  title={Demonstration of disturbance propagation and amplification in car-following situation for enhancement of vehicle platoon system},
  author={Kim, Jinsoo and Jeong, Jinhan and Jhang, Kyung-young and Park, Jahng-hyon},
  booktitle={2015 IEEE Intelligent Vehicles Symposium (IV)},
  pages={999--1005},
  year={2015},
  organization={IEEE}
}

@article{mcnemar1947note,
  title={Note on the sampling error of the difference between correlated proportions or percentages},
  author={McNemar, Quinn},
  journal={Psychometrika},
  volume={12},
  number={2},
  pages={153--157},
  year={1947},
  publisher={Springer}
}

@article{van1997simple,
  title={A simple procedure for the assessment of acceptance of advanced transport telematics},
  author={Van Der Laan, Jinke D and Heino, Adriaan and De Waard, Dick},
  journal={Transportation Research Part C: Emerging Technologies},
  volume={5},
  number={1},
  pages={1--10},
  year={1997},
  publisher={Elsevier}
}

@MISC{Panthera2022,
author = {Van Gaal, Ruud},
title = {Panthera software},
month = April,
year = {2022},
howpublished={\url{https://www.cruden.com/panthera-software/}}
}

@MISC{IPG2020,
author = {IPg Automotive Group},
title = {IPG Carmaker Reference manual version 9.1},
month = September,
year = {2020},
howpublished={\url{https://www.ipg-automotive.com/}}
}

@book{davies1982psychology,
  title={The psychology of vigilance},
  author={Davies, David Roy and Parasuraman, Raja},
  year={1982},
  publisher={Academic Press}
}


\begin{IEEEbiographynophoto}{Klaas O. Koerten} recieved the M.Sc. degree in mechanical engineering from the Delft University of Technology, Delft, The Netherlands, in 2021. He then went on to research hospitality robotics in a collaboration between the Delft University of Technology and the Hotelschool The Hague, The Hague, Netherlands. His research interests include human machine interactions, robotics and the impact of technology on society.
\end{IEEEbiographynophoto}



\begin{IEEEbiographynophoto}{David A. Abbink} received the M.Sc. and Ph.D. degrees in mechanical engineering from the Delft University of Technology, Delft, The Netherlands, in 2002 and 2006, respectively. His research interests include haptics, shared control, human-automation interaction, and his work therein has received continuous funding from Nissan, Boeing, and personal grants from the Dutch Science Foundation. He is currently a Full Professor with the Delft University of Technology, heading the Section of Human-Robot Interaction. He is a senior member of the IEEE.
\end{IEEEbiographynophoto}

\begin{IEEEbiographynophoto}{Arkady Zgonnikov} received the Specialist degree in applied mathematics from Saint-Petersburg State University, Russia, in 2009, and the Ph.D. degree in computer science and engineering from the University of Aizu, Japan, in 2014. He was then an Irish Research Council Postdoctoral Researcher in the Department of Psychology, National University of Ireland Galway, and is currently an assistant professor with Delft University of Technology, The Netherlands. His research interests span the areas of cognitive science, human-robot interaction, and artificial intelligence.
\end{IEEEbiographynophoto}

\end{document}